\documentclass[aps,prd,twocolumn,amsmath,amssymb, showpacs]{revtex4-1}
\usepackage{color}
\usepackage{graphics}
\usepackage{epsfig}
\usepackage{amsmath}
\usepackage{amssymb}
\usepackage{amsthm}
\usepackage[mathscr]{eucal}

\newcommand{\bea}{\begin{eqnarray}}
\newcommand{\eea}{\end{eqnarray}}
\newcommand{\beas}{\begin{eqnarray*}}
\newcommand{\eeas}{\end{eqnarray*}}

\begin{document}

\title{Quark Antiscreening at Strong Magnetic Field and Inverse Magnetic Catalysis}
\author{E. J. Ferrer$^1$, V. de la Incera$^1$, and X. J. Wen$^{1,2}$}
\affiliation{$^1$ Department of Physics, University of Texas at El
Paso, 500 W. University Ave., El Paso, TX 79968, USA\\
$^2$ Institute of Theoretical Physics, Shanxi University, Taiyuan
030006, China}

\begin{abstract}

The dependence of the QCD coupling constant with a strong magnetic field and the implications for the critical temperature of the chiral phase transition are investigated.  It is found that the coupling constant becomes anisotropic in a strong magnetic field and  that the quarks, confined by the field to the lowest Landau level where they pair with antiquarks, produce an antiscreening effect. These results lead to inverse magnetic catalysis, providing a natural explanation for the behavior of the critical temperature in the strong-field region.
\end{abstract}


\maketitle

\section{Introduction}

The study of the QCD phase diagram in the temperature ($T$) - density ($\mu$) plane is a topic that has attracted much attention during many years \cite{QCD-Phases}. The more recent possibility to experimentally reach the high-energy regions where the quark-gluon plasma is realized has activated this research field even more. 

Additionally, the effects of strong magnetic fields in quark matter have also been under active scrutiny for a long time  \cite{CSB}-\cite{Voltage}. At present, such studies have been reactivated by the possibility to reach magnetic field strengths in heavy ion collisions in an energy range which is beyond the intrinsic QCD scale $\Lambda_{QCD}\sim 200$ MeV. 

There are both theoretical and experimental indications that the colliding charged ions can generate very strong magnetic fields, estimated to be of order $eB\sim m^2_\pi (\sim10^{18} G)$ for the top collision, $\surd s_{NN}=$ 200 GeV, in non-central Au-Au impacts at RHIC, and even larger, $eB\sim 15 m^2_\pi (\sim 10^{19} G)$, at the LHC experiments  \cite{LHC}. The magnetic field generated during those collisions is transient. It decays to a tenth of its value in a time scale of order $1/Q_s$ \cite{Magnetic-Decay}, where $Q_s$ is the saturation scale at RHIC \cite{Saturation-Scale}.  On the other hand, a comprehensive comparison between the experiments and a hydrodynamic description of the ultra relativistic collisions done at \cite{QGP} has provided very strong arguments for the creation of a well-developed, thermalized quark-gluon plasma at RHIC just after 0.5 fm/c of the collision, with a significant lifetime of about 5-7 fm/c, and an initial energy density which exceeds the critical value for color deconfinement by at least an order of magnitude. During this time interval the magnetic field is still very near to its maximum strength. Furthermore, during the entire Quark-Gluon plasma (QGP) life-time, the generated magnetic field has been shown to be a slowly varying function of time, i.e. approximately stationary \cite{Magnetic-Decay}. Therefore, it makes sense to explore the effects of the magnetic field on the generated QGP assuming a constant field. 

In a similar fashion, one can also consider that the magnetic field interacting with the QGP is approximately uniform. To understand this, notice that even though the spatial distribution of the magnetic field is globally inhomogeneous, in the central region of the overlapping nuclei, where the QGP is formed, the variation of the magnetic field in the transverse direction is much weaker than the variation of the energy density,  a fact confirmed using the hadron-string-dynamics model \cite{Homogeneous} for $Au-Au$ collisions at $\surd s_{NN}=$ 200 GeV with impact parameter $b=10$ fm.

Based on the above considerations, it is reasonable to assume a uniform and constant magnetic field to qualitatively study the physical characteristics of the chiral phase transition in the magnetized QGP produced in heavy-ion collisions.

It is known that a magnetic field can affect the QCD chiral phase transition. For massless quarks, a magnetic field can catalyze the breaking of chiral symmetry because the attractive interaction between quarks and antiquarks, no matter how weak, is strengthened by the dimensional reduction of the dynamics of the fermions in the lowest Landau level (LLL). This phenomenon, known as magnetic catalysis of chiral symmetry breaking (MC$\chi$SB), leads to the generation of a chiral condensate that modifies the vacuum properties and induces field-dependent dynamical parameters. The MC$\chi$SB has been actively investigated during the last two decades~\cite{severalaspects}-\cite{ferrerincera}.  

In the original studies of magnetic catalysis ~\cite{severalaspects}, the phenomenon was assumed to produce only the conventional  scalar condensate $<\bar{\psi} \psi>$, which in turn gives rise to the generation of a single dynamical parameter: the fermion mass. In recent years, however, it has been found \cite{ferrerincera} - \cite{NJL-B}, that the same universal mechanism of MC$\chi$SB is also responsible for the formation of a magnetic-moment condensate $<\bar{\psi} \Sigma^3 \psi>$, with $\Sigma^3= \frac{i}{2}[\gamma^1,\gamma^2]$ the spin-projection operator along the magnetic field, and the consequential generation of a dynamical anomalous magnetic moment (AMM) for the quasiparticles. In the case of quarks, the existence of a magnetic moment condensate was shown within a Nambu-Jona-Lasinio (NJL) model with interaction channels consistent with the symmetries of QCD in the presence of a magnetic field \cite{NJL-B}. The magnetic-moment condensate produces an increase of the chiral transition temperature.

So far, all the studies of the MC$\chi$SB phenomenon in QCD have led to an increase of both the dynamical mass, (for a review see \cite {Gatto}  and references therein), and the AMM \cite{NJL-B} with the magnetic field. Consequently, the chiral transition temperature $T_c$, which is always of the order of the dynamical mass, has been found to increase with the magnetic field. Nevertheless, this result is in sharp contrast with recent QCD-lattice calculations that showed a decrease of the critical temperature for the chiral/deconfinement transition with the magnetic field \cite{Lattice}, a phenomenon that has been termed "inverse magnetic catalysis" (IMC).

Several attempts to address this disagreement already exist in the literature \cite{Fuku-Hidaka}-\cite {Farias}. In \cite{Fuku-Hidaka}, the authors argued that IMC is already embedded in the NJL approach because the effects of neutral mesons (Goldstone bosons) would suppress the chiral condensate at magnetic fields stronger than the scale of the hadron structure. However, this point of view was later challenged by the results of Ref. \cite{FRG}.  Several authors have argued that the behavior of $T_c$ in lattice QCD should be connected to the dynamics of confinement. For example,  the authors of Ref. \cite{FragaPRD87}  considered the idea that a small number of light quark flavors should decrease the value of the deconfinement critical temperature in the large $N_c$ limit, and extrapolated it to the case with magnetic field. Since they could not explicitly calculate the dependence of the strong coupling with the field, their results were obtained assuming that the expression for $T_c$ would be similar to the case at zero B, but with the coupling replaced by some unknown, positive defined function of B.  On the other hand, the analysis in \cite{Kojo} was based on the proposition that while in the strong field region the chiral condensate grows linearly with the field \cite{Lattice,PLB682}, the dynamical quark mass should be however nearly field-independent in this region. Other efforts \cite{PRD89} introduced the effects of confinement in the NJL model through the Polyakov loop, using the so-called  PNJL model, and argued that one could reproduce the IMC if the lattice data were fitted by making the critical temperature  $T_c$ a parameter of the PNJL model. In \cite{Farias}, IMC was connected to the running of the coupling with the magnetic field. However, the main point of \cite{Farias} was to propose an ansatz for the NJL coupling $G$ that assumed a logarithmic dependence with $B$,  an assumption that is not reliable in the strong-field region, as we will show below.

In the present paper, we adopt the point of view, shared by several authors, that the origin of the IMC should lie in the effects of the magnetic field in the running of the strong coupling. However, our analysis contains two new fundamental elements. On the one hand, we show that in the strong field region ($qB \gg\Lambda^2_{QCD}$), where the infrared dynamics is relevant, the QCD running coupling becomes anisotropic: the color interaction in the directions parallel and transverse to the field is characterized by two different functions of the momentum and the field. On the other hand, we find that the quarks, confined by the field to the LLL, produce antiscreening in the parallel coupling, which is the one entering in the chiral critical temperature. The antiscreening of the LLL quarks is connected, as will become clear below, to the color paramagnetic behavior of the pairs formed by LLL virtual quarks and antiquarks. The antiscreening produced by the LLL pairs increases with the magnetic field because the phase space of the LLL increases with the field, allowing more pairs to be formed. These results naturally lead to IMC and also allow us to identify the physical mechanism behind the behavior of $T_{C_\chi}$ with the field.

As will be shown below, the antiscreening effect of the LLL is apparent in the expression of the color Coulomb potential, because the sign of the contribution coming from the LLL quarks is the same as the sign of the gluon contribution. This is quite different from the situation with no field in the subcritical coupling region where unpaired quarks always produce screening because their contribution to the color Coulomb potential is $\sim - N_f$ and thus enters with a positive sign in the QCD $\beta$-function, in contrast to the gluons, whose contribution to the Coulomb potential is ($\sim N_c$) and thereby negative in $\beta$. As well known, the negative sign of the QCD $\beta$-function is responsible for the color-charge antiscreening which in turn leads to the phenomenon of asymptotic freedom. 


The paper is organized as follow. In Section II, we investigate the running of the strong coupling constant with the magnetic field. We show that at strong magnetic fields the coupling constant becomes anisotropic giving rise to noticeable different interactions in the directions parallel and transverse to the field. Analytical expressions of the parallel and transverse running couplings are obtained, and the profile of the parallel running coupling with the magnetic field for various momenta is shown. In Section III,  the critical temperature of the chiral phase transition is found for a modified NJL model that is in agreement with the symmetries of QCD in a magnetic field and incorporates the running of the coupling with the field in the four-fermion vertex. Taking into account the running of the coupling with the strong field in the infrared region, we find that the critical temperature decreases with the field, in agreement with the findings in lattice QCD. In Section IV, we summarize the main results and discuss the physical origin of the IMC phenomenon.    
 
\section{Anisotropic Coupling Constant}\label{SecII}

 One limitation of the QCD-inspired, effective low-energy NJL-like models where the MC$\chi$SB in QCD has been found, is that the role of the gluons is reduced to multi-fermion point-contact interactions, so the model is not renormalizable and hence does not incorporate the running of the coupling with the scale. Within this framework, magnetically catalyzed parameters like the dynamical mass and the dynamical AMM consistently increase with the magnetic field. Given that the critical temperature $T_{C_{\chi}}$ for the transition to the chirally restored phase is always of the order of the dynamical mass, it follows the same pattern behavior with the field as the mass. 
 
 The NJL approach and lattice QCD are both good to explore the nonperturbative region of QCD, but while the first considers a strong coupling constant independent of the momentum and the field, the second automatically incorporates the effects of confinement and running.  It is then natural to expect that the existence of IMC in lattice QCD and its absence in the conventional NJL model is connected to this main difference between these two approaches. Will IMC emerge if we introduce in the NJL model the field and momentum dependence of the coupling? The answer is yes, as will be shown below. Now, it is important to highlight that in order to explore the magnetic catalysis considering the effect of the field in the running coupling, we must be sure that we work on a region of momenta where the mechanism of MC$\chi$SB is operative, meaning where the fermions remain mostly confined to the LLL. This implies that the average momenta exchanged by particles should be smaller than $\sqrt{qB}$. At the same time, it is known that the most relevant contribution to the magnetically catalyzed parameters come from momenta in the region $m^2_{dyn} < k^2 \ll |qB|$ \cite{Miransky-NPB}, with $m_{dyn}$ the magnetically catalyzed mass. In this infrared region, the QCD expansion may begin to show some well-known problems like infrared poles, lack of convergence, etc., indicating the inconsistency of the standard perturbation series in this region. To consistently explore the running of the coupling in this intermediate infrared region, one needs to incorporate non-perturbative effects in the perturbative expansion of QCD, a method that has been developed over the years in a series of seminal papers \cite{Simonov}.
 
The behavior of the coupling with the magnetic field and the momentum can be extracted from the color Coulomb potential calculated in the presence of a magnetic field. However, to better understand the physical effects of the field on the running of the strong coupling, it is convenient to first recall the case without magnetic field.  

The color Coulomb potential is
\begin{equation}
V(k)=-\frac{4}{3}\frac{4\pi \alpha_s^0(\mu_0)}{k^2+\frac{\alpha_s^0(\mu_0))}{4\pi}(\Pi_g-\Pi_q)}
\label{CP-Pi}
\end{equation}
where $\alpha_s^0(\mu_0)=\frac{g^2(\mu_0)}{4\pi}$ with $\mu_0$ the renormalization energy scale, and $k^2=k_1^2+k_2^2+k_3^2$. The functions $\Pi_g$ and $\Pi_q$ are scalar coefficients associated respectively to the gluon and quark loops contributing to the gluon self-energy that dresses the gluon propagator. In standard perturbation theory they are given by
\begin{equation}
\Pi_g(k)=\frac{11}{3}N_c k^2 \ln\frac{k^2}{\mu_0^2}, \quad  \Pi_q(k)=\frac{2}{3}N_f k^2 \ln\frac{k^2}{\mu_0^2}
\label{Pis}
\end{equation}
respectively.

Using (\ref{Pis}) in (\ref{CP-Pi}), it can be written as
\begin{equation}
V(k)=-\frac{4}{3}\frac{4\pi\alpha_s^0(\mu_0)}{\epsilon(k)k^2}
\label{Coulomb-Potential_0B}
\end{equation} 
where 
\begin{equation}
\epsilon(k)\equiv 1+\frac{\alpha_s^0(\mu_0)}{4\pi}(\frac{11}{3}N_c-\frac{2}{3}N_f)\ln\frac{k^2}{\mu_0^2} 
\label{dielectric constant}
\end{equation} 
can be interpreted as the color electric permittivity in momentum space. As discussed in \cite{Nielsen}, the relativistic invariance of the theory requires that the (color) electric permittivity and the (color) magnetic permeability be connected through the condition
\begin{equation}
\epsilon\mu=1.
\label{emurelation}
\end{equation} 

Considering $k^2 < \mu_0^2$, but far from the pole to avoid the infrared issues of the standard perturbative expansion, we can see from (\ref{Coulomb-Potential_0B}) and (\ref{dielectric constant}) that $\epsilon(k) < 1$ and  $\epsilon(k) \to 1$ when $k \to \mu_0$, so the effective coupling $\alpha(k)=\alpha_s^0(\mu_0)/\epsilon(k)$ decreases with increasing energy scale. This is the characteristic behavior of antiscreening, which leads to asymptotic freedom in QCD. Since (\ref{emurelation}) implies that $\mu=\epsilon^{-1}$, the magnetic permeability exhibits the opposite behavior, so in the same region of momenta one has $\mu (k)> 1$, characteristic of paramagnetism, and  $\mu (k)$ tends to $1$ as $k \to \mu_0$, hence decreasing when the energy scale increases. The connection between color paramagnetism and asymptotic freedom was first highlighted in \cite{Nielsen}.

Notice that the quark and gluon contributions enter in (\ref{dielectric constant}) with opposite signs. Because of this, the quarks tend to screen the color charge, while gluons tend to antiscreen it. As the gluon term wins over the quark's, color charge is antiscreened in QCD.

Let us consider now the color Coulomb potential in the presence of a strong magnetic field. As mentioned above, we are interested in the intermediate infrared region of momenta where the nonperturbative effects of the QCD background cannot be ignored. Fortunately, we can take advantage of existing results \cite{Simonov, Voltage} that have been able to consistently incorporate such nonperturbative effects into the QCD perturbative expansion through the introduction of a nontrivial background of gluon vacuum configurations. Using this background perturbative method (BPM), the gluon loop contribution to the gluon self-energy has been shown to be infrared-finite and equal to 
\begin{equation}
\widetilde{\Pi}_g(k)=\frac{11}{3}N_c k^2 \ln\frac{k^2+M_B^2}{\mu_0^2}, 
\label{Pi_g-nonpertubat}
\end{equation}
where $M_B\approx 1$ GeV is an infrared mass that can be interpreted as the ground-state mass of two gluons connected by the fundamental string, with string tension $\tau=0.18$ GeV$^2$  \cite{Simonov, Voltage}.

A magnetic field affects the color Coulomb potential through quark loops with gluon external legs.  If the field is strong enough to force the quarks to remain in the LLL, the loops of these LLL quarks will lead to a significant anisotropy in the gluon self-energy and hence in the coupling because these loops only contribute to the longitudinal components of the self-energy. The LLL contribution to the gluon self-energy can be found from a calculation very similar to the one done in QED \cite{Batalin} to obtain the LLL  electron contribution to the zero component of the one-loop photon polarization operator,   
\begin{equation}
\frac{\alpha}{4\pi}\Pi^B_{QED} (\textbf{k})=- \frac{2\alpha |eB|}{\pi} \exp \left(\frac{-k_\bot^2}{2|eB|}\right)T\left(\frac{k_3^2}{4m^2}\right),
\label{PiQED}
\end{equation}
where  $k_\bot^2=k_1^2+k_2^2$, $m$ is the renormalized fermion mass, and 
\begin{equation}
T(z)=1-\frac{1}{2\sqrt{z(z+1)}}\ln\frac{\sqrt{1+z}+\sqrt{z}}{\sqrt{1+z}-\sqrt{z}}
\label{T-function}
\end{equation} 
satisfies $0 \le T(z)\le1$, with $T(z) \simeq \frac{2}{3}z$ for $z\ll1$ and $T(z)=1$ for $z\gg1$. 

For the QCD case we just have to replace $\alpha \to \alpha_s^0$ in (\ref{PiQED}), and take into account the difference in the electric charges of different flavors, so that $e \to q_i$ and we have to sum in $i$. The use of the BPM requires in addition to replace $m^2$ by the string tension $\tau$ and use $\alpha_s^0(\mu_0)=12\pi/[11N_c \ln\left((\mu_0^2+M_B^2)/\Lambda_{V}^2\right)]$ \cite{Voltage}. 

For the region of momenta $k_3^2 \ll 4\tau$, we can approximate $T(k^2_3/4\tau)\sim \frac{2}{3}\frac{k^2_3}{4\tau}$ to find
\begin{equation}
\frac{\alpha_s^0}{4\pi}\Pi^B_q(\textbf{k})=k_3^2\frac{\alpha_s^0(\mu_0)}{3\pi}\sum_{i=1}^{N_f}\frac{ |q_iB|}{\tau} \exp \left(\frac{-k_\bot^2}{2|q_iB|}\right)
\label{Piq}
\end{equation}

Using the expressions (\ref{Pi_g-nonpertubat}) and (\ref{Piq}) in (\ref{CP-Pi}), one can readily find the BPM Coulomb potential in a magnetic field \cite{Voltage}, which can be conveniently written to explicitly show the field-induced anisotropy,
\begin{equation}
V(\textbf{k})=\frac{-16\pi}{3}\frac{\alpha_s^0(\mu_0)}{C_\bot k_\bot^2+C_\| k_3^2}.
\label{Coulomb-Potential}
\end{equation}
with 
\begin{equation}
C_\bot=1+\frac{11\alpha_s^0(\mu_0)N_c}{12\pi}\ln\left(\frac{\textbf{k}^2+M_B^2}{\mu_0^2}\right),
\label{POP-Coeff-1}
\end{equation}
and
\begin{equation}
C_\|=C_\bot+\frac{\alpha_s^0(\mu_0)}{3\pi}\sum_{i=1}^{N_f}\frac{ |q_iB|}{\tau} \exp \left(\frac{-k_\bot^2}{2|q_iB|}\right).
\label{POP-Coeff-2}
\end{equation}
 
The consistency of the LLL approximation requires $k^2 \ll q_iB$ and we assume $q_iB \gg \Lambda^2_{QCD}$. Notice that contrary to the case of an applied electric field \cite{Fradkin}, a magnetic field does not change the string tension for a neutral string \cite{Porrati}, which links a quark with an antiquark. 

It should be pointed out that a complete formulation of the infrared behavior of $\alpha_s(k)$ is one of the most challenging problems in QCD. In this sense, the approach followed here, which is based on background perturbation theory and incorporates vacuum nonperturbative configurations in the perturbative expansion \cite{Simonov}, has the advantage that it reproduces the asymptotic freedom of the regular perturbation theory and ensures confinement in the infrared region.  

\begin{figure} \label{Fig1}
\begin{center}
\includegraphics[width=0.40 \textwidth]{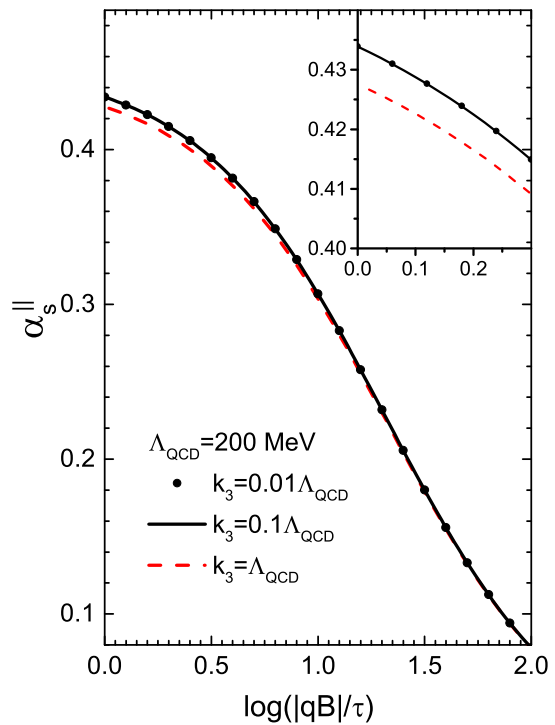}
\caption{\footnotesize (Color online) Coupling constant $\alpha_s^\|$, as a function of the magnetic field at different energy scales. Here we used $\mu_0=1.1$ GeV and $\Lambda_V=0.385$ GeV.} \label{Fig1}
\end{center}
\end{figure}

From (\ref{Coulomb-Potential}), a splitting of the couplings for momenta parallel and transverse to the field follows. They are given respectively by
\begin{equation}
\alpha_{s}^\|(k_3)=\frac{\alpha_s^0(\mu_0)}{C_\| (k_\bot=0)}, \quad \alpha_{s}^\bot(k_\bot)=\frac{\alpha_s^0(\mu_0)}{C_\bot (k_3=0)}
\label{Alpha-1}
\end{equation}

Notice that the quark contribution to $\alpha_{s}^\|(k_3)$ does not depend on the momentum. This means that the LLL quarks do not contribute to the running of $\alpha_{s}^\|(k_3)$ with $k_3$, a fact that has already been pointed out in \cite{Hong}. It is a consequence of the LLL dimensional reduction that gives rise to finite quark loop terms. On the other hand, $\alpha_{s}^\|(k_3)$ runs with the magnetic field showing the typical antiscreening behavior with an energy scale; the larger the field strength, the smaller the coupling, as seen in Fig.\ref{Fig1}. We can also gather from this figure that the decrease of the parallel coupling with the magnetic field is practically unaffected by the infrared energy scale. The curves basically overlap for $k_3=(0.01, 0.1, 1)\Lambda_{QCD}$. In contrast, $\alpha_{s}^\bot(k_\bot)$ does not depend on the magnetic field because it does not receive any contribution from the LLL quarks.

\section{Behavior of $T_c$ with B}\label{SecIII}

As shown in \cite{NJL-B},  the breaking of the rotational symmetry by a uniform magnetic field $B$ induces a separation between longitudinal and transverse fermion modes. This separation leads to the effective splitting of the couplings in the one-gluon exchange interactions on which the NJL models are usually based. This splitting is therefore reflected in the four-fermion couplings of a QCD-inspired NJL model in a magnetic field, and one can use the Fierz identities in a magnetic field \cite{NJL-B}, to show that the NJL Lagrangian in this case should be of the form

\begin{eqnarray} \label{lagrangian}
\mathcal{L}&=&\bar{\psi}i\gamma^{\mu}D_{\mu}\psi+\frac{G}{2}[(\bar{\psi}\psi)^2+(\bar{\psi}i\gamma^5\psi)^2] \nonumber
\\
&+&\frac{G'}{2}[(\bar{\psi}\Sigma^3\psi)^2+(\bar{\psi}i\gamma^5\Sigma^3\psi)^2] \qquad 
\end{eqnarray}
with $D_\mu=\partial_\mu+iqA^{ext}_\mu$, $A^{ext}_\mu=(0,0,Bx_1,0)$,  for a constant and homogenous magnetic field $B$ in the $x_3$-direction,  and $\Sigma^3=\frac{i}{2}[\gamma^1, \gamma^2]$ (see \cite{NJL-B} for details). The couplings $G$ and $G'$ are related to the split gluon-quark vertex couplings  $g_{\|}$ and $g_{\bot}$ through $G=(g_\|^2+g_\bot^2)/2\Lambda^2$, $G'=(g^2_\|-g^2_\bot)/2\Lambda^2$, with $\Lambda$ the energy scale of the effective NJL theory. We can define $G'=\eta G$ with $0\leqslant \eta \leqslant1$. In the strong-field region, $eB/\Lambda^2\sim 1$, and then $\eta \simeq 1$. In this region, all the fermions are confined to the LLL and the only modes contributing to the coupling are the longitudinal ones. Therefore, the separation between longitudinal and parallel modes induced by the contribution of the LLL can be seen as the culprit for the anisotropy manifested in the strong-coupling vertex in the presence of a magnetic field. 

In the mean-field approximation, one can show \cite{NJL-B} that the theory (\ref{lagrangian})  has two separate chiral condensates, $
\overline{\sigma}=-G\langle\bar{\psi}\psi\rangle$ and $\overline{\xi}=-G'\langle\bar{\psi}i\gamma^1 \gamma^2\psi\rangle$, that minimize the free-energy and give rise, in the strong-field region, to a dynamical mass
\begin{equation}\label{sigma}
\overline{\sigma}=\left(\frac{2G\Lambda}{G+G'}\right )\exp{\left[\frac{-2\pi^2}{(G+G')N_cqB}\right]},
\end{equation}
and a dynamical anomalous magnetic moment
\begin{equation}\label{AMM-Solution}
\overline{\xi}=\frac{G'}{G}\overline{\sigma}
\end{equation}
respectively.
The nonperturbative character of the NJL effective-model approach is evident from the way the dynamical parameters $\overline{\sigma}$ and $\overline{\xi}$ depend on the couplings $G$ and  $G'$.

The results (\ref{sigma}) and (\ref{AMM-Solution}) were found using the mean-field approximation. Would they remain valid beyond such an approximation? The reliability of the mean-field approximation to investigate the MC$\chi$SB phenomenon was first addressed in Ref. \cite{Miransky-NPB}, where it was concluded that the contribution of the Nambu-Goldstone fluctuations do not affect the condensation that occurs in the reduced dimension of the LLL quarks, since the Goldstone bosons, being neutral, do not feel the dimensional reduction induced by the magnetic field, hence they are not subject to the consequences of the Mermin-Wagner theorem \cite{MW}. In a recent paper \cite{Fuku-Hidaka}, this conclusion was reconsidered under the reasoning that although the Nambu-Goldstone bosons are neutral, they are composite bosons, formed by a pair of charged fermions each of which is subject to the dimensional reduction produced by the field.  According to this idea, the dimensional reduction in the dynamics of the Goldstone bosons would lead to a decrease of the dynamical mass with the field, in clear opposition to the mean-field MC$\chi$SB results. The authors of \cite{Fuku-Hidaka} called this new mechanism Magnetic Inhibition. However, a subsequent investigation \cite{FRG}, based on the functional renormalization group approach- a powerful nonperturbative method to go beyond the mean-field approximation by fully taking into account thermal and quantum fluctuations- redeemed the validity of the MC$\chi$SB mean-field solution. As explained in \cite{FRG}, the problem in the results found in \cite{Fuku-Hidaka} was that they were obtained ignoring the impact of the anisotropic fluctuations of the neutral Goldstone fields at finite temperature. Hence, the mean-field approximation has been shown to be reliable and the results of the MC$\chi$SB within this approximation have been proved to be robust.

Chiral symmetry restoration in NJL models with MC$\chi$SB typically occurs as a second order transition at a critical temperature $T_{C_\chi} \simeq m_d \ll \sqrt{qB}$ with $m_d$ the dynamical mass of the model considered. The critical temperature $T_{C_\chi}$ can be found from
\begin{equation}\label{Critical-Temperature}
\frac{\partial^2\Omega_0^{T_{C_\chi}}}{\partial \overline{\sigma}^2}|_{\overline{\sigma}=\overline{\xi}=0}=0,
\end{equation}
where the one-loop thermodynamic potential in the strong-field limit corresponding to the Lagrangian density (\ref{lagrangian}) is given by \cite{NJL-B}
\begin{eqnarray}\label{Thermo-Potential-2}
\Omega_0^T(\overline{\sigma},\overline{\xi})&=&-N_cqB\int_{0}^\Lambda \frac{dp_3}{2\pi^2}\left[ \varepsilon +\frac{2}{\beta}\ln \left(1+e^{-\beta\varepsilon}\right)\right]\nonumber
\\
&+&\frac{\overline{\sigma}^2}{2G}+\frac{\overline{\xi}^2}{2G'},
\end{eqnarray}
with $\varepsilon^2=p_3^2+(\overline{\sigma}+\overline{\xi})^2$. 

The critical temperature obtained from (\ref{Critical-Temperature})-(\ref{Thermo-Potential-2}) is then
\begin{equation}\label{Critical-Temperature-3}
T_{C_\chi}=1.16\sqrt{qB} \exp \left [\frac{-2\pi^2}{(G+G')N_cqB}\right ].
\end{equation}
The same result is obtained if the derivative in (\ref{Critical-Temperature}) is taken instead with respect to $\overline{\xi}$. This is a consequence of the proportionality between $\overline{\sigma}$ and $\overline{\xi}$, given in Eq. (\ref{AMM-Solution}), which implies that the two condensates evaporate at the same critical temperature.

In the conventional NJL approach, the couplings  $G$ and $G'$ are constants independent of the external conditions, therefore the critical temperature (\ref{Critical-Temperature-3}) increases with increasing $B$.  This is what happens in all chiral-model calculations (for a recent review see \cite{Gatto}). 

However, one can consider an effective NJL model on which we incorporate the effects of the magnetic field in the coupling by using the relation $G=4\pi\alpha_s/\Lambda^2$ and the results for the strong coupling found in Section \ref{SecII}.  Since in the presence of a strong magnetic field, the effective coupling entering in the dynamical mass is actually $G+G'$, the corresponding relation is  then
\begin{equation}\label{Coupling-Constants}
G+G'=\frac{g^2_\|}{\Lambda^2}=\frac{4\pi\alpha^\|_s}{qB},
\end{equation}
where we assumed that all the quarks are in the LLL, so that $\eta\simeq1$, $\Lambda \simeq \sqrt{qB}$, and the momentum transfer between quarks and gluons is effectively driven by the longitudinal modes (this is why in (\ref{Coupling-Constants}) we write $\alpha^\|_s$). In terms of $\alpha^\|_s$, the critical temperature (\ref{Critical-Temperature-3}) can be written as
\begin{equation}\label{Critical-Temperature-4}
T_{C_\chi}=1.16\sqrt{qB} \exp \left [\frac{-\pi}{2N_c \alpha^\|_s} \right ]
\end{equation}

Taking into account the dependence of  $\alpha^\|_s$ with the field shown in Fig. \ref{Fig1} we can readily obtain the profile of the critical temperature with the magnetic field. This is shown in Fig. \ref{Fig2}, which clearly exhibits an IMC behavior, in qualitative agreement with the results of lattice QCD \cite{Lattice}. Here we took $\alpha^\|_s$ from Eq. (\ref{Alpha-1}) for $N_c=3$ and considered several energy scales $k_3 \le \Lambda_{QCD}$. As in Fig. \ref{Fig1}, we used the string tension $\tau=0.18$ GeV$^2$ \cite{Simonov, Voltage} to normalize the magnetic field in the plot. The strong-field approximation $T\ll\sqrt{qB}$ used here allows to neglect any T-dependence in $\alpha^\|_s$. Notice that $T_{C_\chi}$ is consistent with this approximation.

\begin{figure} \label{Fig2}
\begin{center}
\includegraphics[width=0.40 \textwidth]{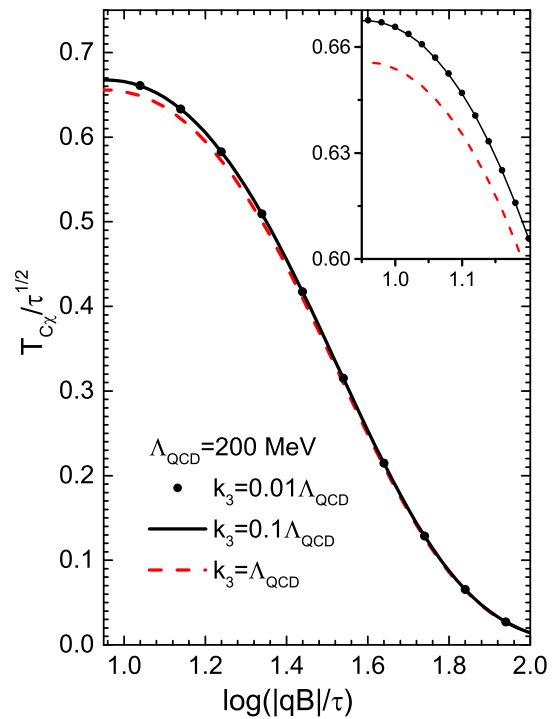}
\caption{\footnotesize (Color online)
Critical temperature for chiral symmetry restoration as a function of the magnetic field at different energy scales. Here we used the string tension $\tau=0.18$ GeV$^2$ as a normalization scale for the field and temperature.} \label{Fig2}
\end{center}
\end{figure}

\section{Concluding Remarks}

In this paper we studied the effects of a strong magnetic field in the coupling of quarks and gluons in QCD and used it to investigate the behavior of the chiral transition temperature with the field. To accomplish this goal, we considered the phenomenon of MC$\chi$SB in a modified NJL model that incorporates the effects of the field in the running coupling. To explore the coupling constant behavior in the region of momenta relevant for the MC$\chi$SB mechanism, we extracted the coupling from the color Coulomb potential calculated within the BPM approach. The BPM considers the QCD perturbative series in a background of vacuum nonperturbative configurations \cite{Simonov}. It allows to consistently investigate the behavior of the coupling in the intermediate region of momenta where the standard perturbative series shows several inconsistencies. The BPM produces infrared regulators, avoid infrared renormalons, and have the advantage that it can describe both confinement and asymptotic freedom \cite{Simonov}.  

Using the BMP color Coulomb potential in a magnetic field, we found that in the strong-field region the coupling of quarks and gluons becomes anisotropic with respect to the directions parallel and transverse to the magnetic field. The transverse coupling does not get contributions from quark loops, so it does not change with the magnetic field, but the parallel coupling gets contributions from the quarks and decreases with the field thanks to the contribution of the LLL quarks. 

Recall that in the standard perturbative expansion the running of the coupling constant in the absence of a magnetic field is given by
\begin{equation}\label{Coup-Const-Vacuum}
\alpha_{s}(k)=\frac{\alpha_s^0(\mu_0)}{1+\frac{\alpha_s^0(\mu_0)}{4\pi}(\frac{11}{3}N_c-\frac{2}{3}N_f) \ln (k^2/\mu_0^2)}.
\end{equation}
The contribution of the quarks enters through the term $-\frac{2}{3}N_f\ln (k^2/\mu^2)$, which tends to increase (decrease) the strength of the coupling at large (small) energies. This is the usual effect of unpaired quarks on the strong coupling at small (large) distances, very similar to the screening effect of the electric charge by charged fermions. In contrast, the gluon part, proportional to $N_c \ln (k^2/\mu^2)$, enters with a sign opposite to the quark contribution, so the tendency from this term is to decrease (increase) the coupling at large (small) energies, and hence displays antiscreening at large distances.  

In a strong magnetic field, the quarks are confined to their LLL where they pair with antiquark via the MC$\chi$SB mechanism. Contrary to what happens in the absence of a magnetic field, the LLL quarks contribute to $\alpha^\|_s$ with a positive sign (see Eqs. (\ref{POP-Coeff-1})-(\ref{Alpha-1})). Because of this, they produce antiscreening just as the gluons. The physical mechanism behind this radical change of the quarks' effect on the color charge can be understood as follows. In the infrared region all the virtual quarks and antiquarks are paired via the MC$\chi$SB. The pairs of LLL quarks and antiquarks form magnetic dipoles that align themselves with the external magnetic field. This means that the LLL behaves as an electromagnetic paramagnet. The paired LLL quarks are not only magnetic dipoles, but because the quark and antiquark in the pair have opposite spins and opposite color charges, they are also color-magnetic dipoles. Once they become aligned with the magnetic field, they inevitably produce a net alignment of their chromomagnetic moment too. Hence, the LLL quarks also behave as a color paramagnet. In a relativistic theory paramagnetism implies antiscreening and viceversa \cite{Nielsen}. Hence, the color paramagnetism of the LLL pairs gives rise to color antiscreening. Notice that the quark-antiquark that pair at the LLL also form color electric dipoles that will orient, as gluons do, to antiscreen a test color charge in the parallel direction. The antiscreening effect of the magnetic field on the parallel coupling constant increases with the field because the larger the field, the larger the density of states of the LLL,  so more dipoles can be formed.

Looking at the way $\alpha^\|_s$ enters in the critical temperature (\ref{Critical-Temperature-4}), it is evident that only if $\alpha^\|_s$ decreases with the magnetic field, can $T_c$ also decrease with the field and hence exhibit the same behavior found in lattice QCD. On the other hand, the magnetic field can only enter in the running coupling through the quark loops. Therefore, the inverse magnetic catalysis can be directly linked to the antiscreening effect of the paired quarks in the LLL and their alignment in the external magnetic field.  

Finally, we call the reader's attention to the fact that the behavior of $\alpha^\|_s$ with the magnetic field would be even sharper if, as recently found in lattice QCD \cite{lattice-Btension}, the string tension would also decrease with the field. A sharper drop in $\alpha^\|_s$ would lead in turn to a sharper decreasing of $T_{C_\chi}$ with $B$, meaning a stronger inverse magnetic catalysis effect. An interesting pending task is to reconcile the results obtained for the string tension in lattice QCD \cite{lattice-Btension}, which those obtained in string theory,  where $\tau$ is independent of the magnetic field \cite{Porrati}.

{\bf Acknowledgments:} The authors thank Dimitri Kharzeev and Massimo D'Elia for insightful discussions.  The work of E. J. Ferrer and V. de la Incera has been supported by DOE Nuclear Theory grant DE-FG02-07ER41458. X. J. Wen thanks the hospitality of University of Texas at El Paso where this work was done and the support of the National Natural Science Foundation of China Grants No. 11475110.

\end{document}